\documentclass[11pt]{article}

\usepackage{amsfonts,amsthm}
\usepackage{amsmath}  
\usepackage{latexsym}
\usepackage{amssymb}
\usepackage{mathrsfs}
\usepackage{tikz}

\newtheorem{thm}{Theorem}[section]

\newtheorem{corollary}{Corollary}[section]

\newcommand{\lda}{\lambda}
\newcommand{\HH}{\mathbb{H}}
\newcommand{\RR}{\mathbb{R}}
\newcommand{\CC}{\mathbb{C}}

\newcommand{\MM}{\mathbb{M}}

\newcommand{\cb}{\mathfrak{b}}
\newcommand{\ct}{\mathfrak{t}}
\newcommand{\1}{\mbox{\hspace{.0em}1\hspace{-.24em}I}}

\begin{document}

\begin{center}
\textbf{\Large From Hamiltonian to zero curvature formulation for classical integrable boundary conditions}\\[3ex]
\large{Jean Avan$^a$, Vincent Caudrelier$^b$, Nicolas Cramp\'e$^c$}\\[3ex]

$^a$Laboratoire de Physique Th\'eorique et Mod\'elisation (CNRS UMR 8089),\\
Universit\'e de Cergy-Pontoise, F-95302 Cergy-Pontoise, France\\[3ex]

$^b$ School of Mathematics, University of Leeds, LS2 9JT, UK  \\[3ex]

$^c$Laboratoire Charles Coulomb (L2C), UMR 5221 CNRS-Universit{\'e} de Montpellier,\\
Montpellier, F-France.\\[3ex]
 
\end{center}

\vspace{0.5cm}

\centerline{\bf Abstract}

We reconcile the Hamiltonian formalism and the zero curvature representation in the approach to integrable boundary conditions for a classical integrable system in $1+1$ space-time dimensions. We start from
an ultralocal Poisson algebra involving a Lax matrix and two (dynamical) boundary matrices. Sklyanin's formula for the double-row transfer matrix is used to derive Hamilton's equations of motion for both the Lax matrix {\bf and} the boundary matrices in the form of zero curvature equations. A key ingredient of the method is a boundary version of the Semenov-Tian-Shansky formula for the generating function of the time-part of a Lax pair. The procedure is illustrated on the finite Toda chain for which we derive Lax pairs of size $2\times 2$ for previously known Hamiltonians of type $BC_N$ and $D_N$ corresponding to constant and dynamical boundary matrices respectively.

\section*{Introduction}

Since the seminal paper \cite{Sk} on integrable boundary conditions for classical systems\footnote{In this paper, we only consider classical systems and hence the even larger literature on quantum integrable boundary conditions, based on the seminal paper also by Sklyanin, will not be mentioned.} (see also \cite{Che}) , a large amount of work and progress has been 
made in describing models on the half-line or the interval\footnote{We include both models with a discrete or continuous space variable under this terminology. For the former, fields depends on an integer $j$ and time $t$. For the latter, they depend on a real variable $x$ and time $t$.}. It is important to realise that \cite{Sk} contains two distinct but related points of view on the question of integrability in the presence of boundary conditions, each of which having evolved into rather separate areas. 

The first point of view is purely Hamiltonian and rests upon two cornerstones of integrability: a (quadratic/linear) Poisson algebra satisfied by a Lax matrix $\ell(j,\lda)$/$\ell(x,\lda)$ (discrete/continuous case), and the classical reflection equation for (constant) matrices $k^\pm(\lda)$, $\lda$ being the spectral parameter. Both structures involve a fundamental object: the classical $r$-matrix, encoding the class of systems one is working with. The celebrated result is that, given the monodromy matrix $L(\lda)$ constructed from the $\ell$'s and solutions $k^\pm(\lda)$ of the reflection equation, the quantity $\cb(\lda)=tr_a\big(\ k^+_a(\lambda) \ L_a(\lambda) \  k^-_a(\lambda)\  L_a(-\lambda)^{-1}\ \big)$ Poisson commutes with itself for different values of the spectral parameter. One therefore uses a Hamiltonian that can be extracted from $\cb(\lda)$ to generate the time evolution on the fields and derive Hamilton's equations of motion. This aspect has been used to produce expressions for Hamiltonians with integrable boundary terms, see e.g. \cite{Sk,Mc} and \cite{KT,BD} for examples of the so-called dynamical boundary case.

The second point of view focuses on the Lax {\bf pair} and the associated auxiliary spectral problem, and attempts to provide what we could call a zero curvature description of an integrable systems with boundaries. From this point of view, the central equation advocated in \cite{Sk} involves the matrices $k^\pm(\lda)$ and the second matrix of the Lax pair for the bulk equations of motion, denoted $M(j,\lda)$/$M(x,\lda)$. It reads
\begin{equation}
\label{relation_KM}
k^\pm(\lda)M(x^\pm,\pm\lda)=M(x^\pm,\mp\lda)k^\pm(\lda)\,,
\end{equation}
where $x^\pm$ denotes the positions of the two boundaries defining the interval.
This formed the basis of a line of work initiated by Habibullin \cite{H} which led eventually to the notion of nonlinear mirror image method for tackling the Inverse Scattering Method on the half-line, see e.g. \cite{BT1,BT2,T,BH,CZ1,CZ2}. Note also that relation \eqref{relation_KM} has been revived more recently within the framework of the so-called Unified Transform \cite{Fok1} where it is used to define the notion of linearizable boundary conditions.

Having a Hamiltonian and a zero-curvature point of view in classical integrable systems is far from being an issue and has long been identified as one of the crucial aspect of the theory: there is a natural connection between the two pictures and one can speak of Liouville integrability for a PDE. The central object capturing this connection is the classical $r$-matrix \cite{Skly,STS1}. Using the $r$-matrix presentation, one can derive the Semenov-Tian-Shansky formula which provides the time component of the Lax pair, see e.g. \cite{FT} p.199. Thus, the zero-curvature representation of the theory at hand is a consequence of the Hamiltonian formulation. 

However, in the case with boundaries, the state of affairs is rather unsatisfactory for the following two reasons:

\begin{enumerate}
	\item Considering for instance the NLS equation, as in \cite{Sk}, one finds that relation \eqref{relation_KM} restricts the matrices $k^\pm(\lda)$ to be diagonal, of the form $\lda\sigma_3+i\theta^\pm$, $\theta^\pm\in\RR$, producing the well-known Robin boundary conditions. However, in the rational case, the reflection equation allows for the general solution (up to a multiplicative function of $\lda$)
	\begin{equation}
	k(\lda)=\lda\,A+\theta\,\1_2\,,~~A\in\mathsf{sl}(2,\CC)\,.
	\end{equation}
	In particular, even in the reduced case producing NLS, one can have a solution with nonzero off-diagonal elements. How do we resolve this discrepancy? 
	
	\item In the boundary case, a systematic connection between the Hamiltonian approach and the zero-curvature presentation is not available yet in full form, despite important results in \cite{AD1,AD2} where the boundary version of the Semenov-Tian-Shansky formula was derived. This gap in the theory has been described under various forms in the literature, with no satisfactory answer to the best of our knowledge. For instance, in \cite{Cor}, the author reviews some results \cite{BCDR,BCR} which took the view of modifying in an ad hoc fashion the bulk Lax pair in order to accommodate integrable boundary conditions. The dynamical generalisation of \eqref{relation_KM} appears as a consistency condition in this approach. It is used to find admissible, time-independent, $k$ matrices. To ensure that the approach is consistent with the Hamiltonian point of view, it is checked {\it a posteriori} that these solutions are also solutions of the reflection equation. That they do is described as ``remarkable'' as the $k$ matrices are obtained independently of the $r$-matrix from this point of view.  
\end{enumerate}

We have found that reconciling the mismatch in the first point requires the development of the systematic connection raised in point two, in turn providing an answer to the seemingly remarkable connection between the $r$ and $k$ matrices. Analysing the mismatch in the first point also leads us naturally to considering dynamical $k$ matrices even when one starts with solutions of the (non-dynamical) reflection equation originally considered in \cite{Sk}. It is the purpose of this paper to present the general framework 
that addresses the two aforementioned problems. We treat both the discrete and continuous cases as there is no conceptual difference between the two. All the details are given in the discrete case and the results go over to the continuous case with the standard appropriate technical changes. Our main result is Theorem \ref{th_ZC}, and its corollary, which establish the zero-curvature presentation of the equations of motion for both the bulk and boundary fields from the Hamiltonian formalism, in the dynamical boundary case.

The paper is organised as follows. In Section \ref{Sec1}, we first recall, in the discrete case, known constructions that provide the link between the Hamiltonian and zero curvature presentations via the $r$-matrix formalism. We then show how to modify this to accommodate $k$ matrices describing the presence of boundaries and obtain our main result Theorem \ref{th_ZC}. Section \ref{Sec2} contains the analogous formulas and results in the continuous case, without details. The last section contains an example of our constructions: the Toda chain. 
We show that we can {\bf deduce} the Lax pair and zero curvature representation for the $BC_N$ type Toda chain with open boundary conditions which, to the best of our knowledge, was given {\bf ad hoc} in \cite{kS}. We also derive dynamical boundary conditions for the Toda chain which can be matched to boundary conditions found in \cite{AH} by a completely different method. This corresponds to the $D_N$ Toda chain whose Hamiltonian was discussed in \cite{KT} and for which we obtain a Lax pair. 

\section{Hamiltonian and zero-curvature presentations for models with open boundaries: Discrete case}\label{Sec1}

\subsection{Single-row formalism}

We consider discrete models described by a Lax matrix $\ell(j,\lambda)$ and equipped with a Poisson structure, such that the following 
ultralocal Poisson algebra relation holds
\begin{equation}
\label{rll}
 \{\ell_a(j,\lambda)\ ,\ \ell_b(k,\mu)\}=\delta_{jk}\left[ r_{ab}(\lambda-\mu)\ , \ \ell_a(j,\lambda) \ell_b(k,\mu) \right]\,.
\end{equation}
Here we use the so-called auxiliary space notation meaning\footnote{Note that for our purposes, it is enough to work directly in $End(\CC^2\otimes\CC^2)$ although in general, objects like $\ell$ or $r$ live in more abstract structures.}
\begin{equation}
\ell_a(j,\lambda)=\ell(j,\lambda)\otimes \1_2\,,~~\ell_b(j,\lambda)=\1_2\otimes \ell(j,\lambda)\,,~~\{\ell_a(j,\lambda)\ ,\ \ell_b(k,\mu)\}=
\{\ell_{ij}(j,\lambda)\ ,\ \ell_{kl}(k,\mu)\}\,E_{ij}\otimes E_{kl}
\end{equation}
where $\1_2=\delta_{ij}E_{ij}$ is the identity matrix and $\ell=\ell_{ij}E_{ij}$ is a $2\times 2$ matrix (see Section \ref{Sec3} for an example).
The $r$-matrix is assumed to be skew-symmetric $r_{ab}(\lda)=-r_{ba}(-\lda)$ and satisfies the classical Yang-Baxter equation
\begin{eqnarray}
 [\ r_{ac}(\lda-\nu)\ ,\ r_{bc}(\mu-\nu)\ ]+ [\ r_{ab}(\lda-\mu)\ ,\ r_{ac}(\lda-\nu)\ ]+ [\ r_{ab}(\lda-\mu)\ ,\ r_{bc}(\mu-\nu)\ ]=0\,.
\end{eqnarray}
One defines the single-row monodromy matrix by
\begin{equation}
 L(\lambda)=\ell(N,\lambda)\ell(N-1,\lambda)\dots \ell(1,\lambda)\,,
\end{equation}
and the so-called single-row transfer matrix\footnote{The terminology transfer ``matrix'' comes from the quantum setting whereas here, it is clear that $\ct(\lda)$ is simply a scalar.} by taking the trace of $L(\lda)$
\begin{equation}
\ct( \lambda)=tr\, L(\lambda)
\end{equation}
Then the following holds
\begin{eqnarray}
\label{rt}
&& \{L_a(\lambda), L_b(\mu)\}=\left[ r_{ab}(\lambda-\mu)\ , \ L_a(\lambda) \ L_b(\mu) \right]\,,\\
&& \{\ct( \lambda) , \ct(\mu)\}=0\;.
\end{eqnarray}
The second relation allows us to take $\ct(\lambda)$ as the generating function in $\lambda$ of Poisson commuting Hamiltonians $H^{(n)}$.
The Hamiltonian $H$ of the system of interest is expressed in terms of the $H^{(n)}$ and we associate to it an evolution time $t$ according to 
\begin{equation}
 \partial_{t}\ \cdot=\{ H ,\ \cdot \}
\end{equation}
In particular, Hamilton's equations for the fields contained in $\ell(j,\lda)$ are given by
\begin{equation}
\label{eq_Hamilton}
\partial_{t}\ \ell(j,\lda)=\{ H ,\ \ell(j,\lda) \}\,.
\end{equation}
It is a fundamental result that these equations have a (discrete) zero curvature representation. 
Let us define the partial monodromy for $n\geq m$
\begin{equation}
L_a(n,m,\lambda)=\ell_a(n,\lambda)\ell_a(n-1,\lambda)\dots \ell_a(m,\lambda)\;.
\end{equation}
We use the convention $L(n-1,n,\lambda)=1$ and obviously one gets $L(N,1,\lambda)=L(\lambda)$. Following \cite{STS1,Fad}, one defines
\begin{equation}
 M_b(j,\lambda,\mu)=tr_a\big(\ L_a(N,j,\lambda)\ r_{ab}(\lambda-\mu)\ L_a(j-1,1,\lambda)\ \big)
\end{equation}
Here, the notation $tr_a$ means that we take the trace only over the first copy of $\CC^2$.
Then, using relation \eqref{rll}, one finds
\begin{equation}
 \{ \ct(\lambda),\ell(j,\mu) \}=M(j+1,\lambda,\mu)\ell(j,\mu) - \ell(j,\mu) M(j,\lambda,\mu)
\end{equation}
$M(j,\lambda,\mu)$ is the generating function in $\lambda$ of the matrices $M^{(n)}(j,\mu)$ associated to the Hamiltonians  $H^{(n)}$. Denoting by $M(j,\mu)$ the matrix associated to $H$ and its 
corresponding time $t$, one finds that \eqref{eq_Hamilton} has the discrete zero curvature representation
 \begin{equation}
  \partial_{t}\ell(j,\mu)=M(j+1,\mu)\ell(j,\mu) - \ell(j,\mu) M(j,\mu)\;.
 \end{equation}
 In other words, the pair $(\ell(j,\mu), M(j,\mu))$ is a Lax pair for the system under consideration.

 \subsection{Double-row formalism for models with open boundaries}
 In this section, we show how to obtain the boundary version of the above connection between Hamiltonian and zero curvature representation. 
 This is our main result, Theorem \ref{th_ZC} and Corollary \ref{coro_ZC}.
 Let us now assume that the bulk and boundary fields contained in the matrices $\ell(j,\lda)$ and $k^\pm(\lda)$ satisfy the following (ultralocal) boundary Poisson algebra.
\begin{eqnarray}
 \{\ell_a(j,\lambda)\ ,\ \ell_b(k,\mu)\}&=&\delta_{jk}\left[ r_{ab}(\lambda-\mu)\ , \ \ell_a(j,\lambda) \ell_b(k,\mu) \right]\label{rll2}\,,\\
   \{k_a^-(\lambda)\ ,\ k_b^-(\mu) \}&=&r_{ab}(\lambda-\mu)k_a^-(\lambda) k_b^-(\mu) -   k_a^-(\lambda)   \ k_b^-(\mu) r_{ba}(\lambda-\mu)\nonumber\\
   \label{Kdyn1}
   &&+k_a^-(\lambda)r_{ba}(\lambda+\mu) k_b^-(\mu)-k_b^-(\mu)r_{ab}(\lambda+\mu) k_a^-(\lambda)\label{rkk1}\,,\\
  \{k_a^+(\lambda)\ ,\ k_b^+(\mu) \}&=&r_{ba}(\lambda-\mu)k_a^+(\lambda) k_b^+(\mu) -   k_a^+(\lambda)   \ k_b^+(\mu) r_{ab}(\lambda-\mu)\nonumber\\
  \label{Kdyn2}
   &&+k_a^+(\lambda)r_{ab}(\lambda+\mu) k_b^+(\mu)-k_b^+(\mu)r_{ba}(\lambda+\mu) k_a^+(\lambda)\,,\label{rkk2}\\
 \{k^-_a(\lambda)\ ,\ k^+_b(\mu) \}&=&0\label{kloc2}\,,\\
 \{k^\pm_a(\lambda)\ ,\ \ell_b(j,\mu)\}&=&0\;.\label{kloc3}
\end{eqnarray}
Eqs \eqref{Kdyn1}-\eqref{Kdyn2} have the form of the classical limit of the reflection algebras appearing in \cite{Skquant}. 
They are examples of the general quadratic Poisson algebras introduced in \cite{FrMa}. More precisely they can be characterized as dynamically-generalized ``soliton-preserving'' conditions.

Note that although we keep the same notation for the Poisson bracket as in the previous section, we are in fact dealing with an enlarged Poisson manifold also containing the fields in the matrices $k^\pm(\lda)$. 
The double-row transfer matrix is defined by 
\begin{eqnarray}\label{drtr}
 \cb(\lambda)=tr_a\big(\ k^+_a(\lambda) \ L_a(\lambda) \  k^-_a(\lambda)\  L_a(-\lambda)^{-1}\ \big)
\end{eqnarray}
and satisfies, using \eqref{rll2}-\eqref{kloc3},
\begin{eqnarray}
 \{ \cb(\lambda)\ ,\ \cb(\mu) \}=0\;.
\end{eqnarray}

Similarly to $\ct(\lda)$ in the single-row case, here $\cb(\lambda)$ is the generating function in $\lambda$ of Poisson commuting Hamiltonians $\HH^{(n)}$. In general, the Hamiltonian of the system of interest is expressed in terms of the $\HH^{(n)}$'s and we associate a ``double-row evolution time'' $T$ to it according to 
\begin{equation}
 \partial_{T}\ \cdot=\{\HH\ ,\ \cdot \}\,.
\end{equation}
The full set of Hamilton's equations of motions for the bulk and boundary fields is now
\begin{eqnarray}
\partial_{T}\ \ell(j,\lda)&=&\{\HH ,\ \ell(j,\lda) \}\,,\\
\partial_{T}\ k^\pm(\lda)&=&\{\HH ,\ k^\pm(\lda) \}\,.
\end{eqnarray}

We are now in a position to show that these equations have a zero curvature representation. This involves using a formula that allows 
one to effectively compute algorithmically the second matrix of the Lax pair involved in the zero curvature. 
Following the construction of \cite{AD1} which we generalise to the present case of dynamical $k$ matrices, we define, for $j=1,2,\dots,N+1$, 
\begin{eqnarray}\label{eq:Mad}
 \MM_b(j,\lambda,\mu)&=&tr_a\big(\ k^+_a(\lambda) \ L_a(N,j,\lambda)\  r_{ab}(\lambda-\mu)\ L_a(j-1,1,\lambda) \ k^-_a(\lambda)\ L_a(-\lambda)^{-1}\ \big)\nonumber\\
 &+&tr_a\big(\ k^+_a(\lambda)  \ L_a(\lambda)\  \ k^-_a(\lambda)\ L_a(j-1,1,-\lambda)^{-1}\ r_{ba}(\lambda+\mu)\ L_a(N,j,-\lambda)^{-1}\ \big)
\end{eqnarray}
Then we prove
\begin{thm}\label{th_ZC}
 \begin{eqnarray}
 \{ \cb(\lambda)\ , \ \ell_b(j,\mu) \}&=&\MM_b(j+1,\lambda,\mu)\ \ell_b(j,\mu) - \ell_b(j,\mu)\ \MM_b(j,\lambda,\mu)\\
 \{ \cb(\lambda)\ ,\ k_b^-(\mu) \}&=& \MM_b(1,\lambda,\mu)\ k_b^-(\mu) - k_b^-(\mu)\ \MM_b(1,\lambda,-\mu)\\
  \{ \cb(\lambda)\ ,\ k_b^+(\mu) \}&=&\MM_b(N+1,\lambda,-\mu)\ k_b^+(\mu) - k_b^+(\mu)\ \MM_b(N+1,\lambda,\mu)
\end{eqnarray}
\end{thm}
\proof 
 \begin{eqnarray*}
\{ \cb(\lambda)\ , \ \ell_b(j,\mu) \}&=&tr_a\big(\ k^+_a(\lambda) \ L_a(N,j+1,\lambda)
\{ \ell_a(j,\lambda)\ , \ \ell_b(j,\mu) \}L_a(j-1,1,\lambda) \  k^-_a(\lambda)\  L_a(-\lambda)^{-1}\ \big)\\
&&+tr_a\big(\ k^+_a(\lambda) \ L_a(\lambda)  \  k^-_a(\lambda)\  L_a(j-1,1,\lambda)^{-1} 
\{ \ell_a(j,-\lambda)^{-1}\ , \ \ell_b(j,\mu) \}L_a(N,j+1,\lambda)^{-1}\ \big)\,.
\end{eqnarray*}
Upon inserting 
\begin{eqnarray}
 \{\ell_a(j,\lambda)\ ,\ \ell_b(j,\mu)\}&=&\left[ r_{ab}(\lambda-\mu)\ , \ \ell_a(j,\lambda) \ell_b(j,\mu) \right]\,,\\
\{\ell_a(j,-\lambda)^{-1}\ ,\ \ell_b(j,\mu)\}&=&-\ell_a(j,-\lambda)^{-1}r_{ab}(-\lambda-\mu) \ell_b(j,\mu)+\ell_b(j,\mu)r_{ab}(-\lambda-\mu)\ell_a(j,-\lambda)^{-1}
\end{eqnarray}
and using $r_{ab}(-\lambda-\mu)=-r_{ba}(\lambda+\mu)$, the right-hand-side can be rearranged into $\MM_b(j+1,\lambda,\mu)\ \ell_b(j,\mu) - \ell_b(j,\mu)\ \MM_b(j,\lambda,\mu)$ as required. Note that we have also used the following property of the partial trace, for any three matrices $A$, $B$, $C$,
\begin{eqnarray}
tr_a(A_{ab}\,B_b\,C_a)=tr_a(A_{ab}\,C_a)\,B_b\,,~~tr_a(B_b\,A_{ab}\,C_a)=B_b\,tr_a(A_{ab}\,C_a)\,.
\end{eqnarray}
Similarly,
\begin{eqnarray*}
\{ \cb(\lambda)\ ,\ k_b^-(\mu) \}&=& tr_a\big(\ k^+_a(\lambda) \ L_a(\lambda) \  \{k_a^-(\lambda)\ ,\ k_b^-(\mu) \}\  L_a(-\lambda)^{-1}\ \big)\\
   &=& tr_a\big(\ k^+_a(\lambda) \ L_a(\lambda) \ 
(r_{ab}(\lambda-\mu)k_a^-(\lambda)
+k_a^-(\lambda)r_{ba}(\lambda+\mu)) 
    \  L_a(-\lambda)^{-1}\ \big)k_b^-(\mu)\\
&&- k_b^-(\mu)\ tr_a\big(\ k^+_a(\lambda) \ L_a(\lambda) \ 
        (k_a^-(\lambda) r_{ba}(\lambda-\mu)
    +r_{ab}(\lambda+\mu) k_a^-(\lambda))   
    \  L_a(-\lambda)^{-1}\ \big)\\
&=&\MM_b(1,\lambda,\mu)\ k_b^-(\mu) - k_b^-(\mu)\ \MM_b(1,\lambda,-\mu)\\
\end{eqnarray*}
The proof for $\{ \cb(\lambda)\ ,\ k_b^+(\mu) \}$ is similar.
\endproof

As a direct corollary, if we denote $\MM(j,\mu)$ the matrix extracted from $\MM(j,\lambda,\mu)$ consistently with our extraction of 
$\HH$ from $\cb(\lda)$, we obtain the zero curvature representation of the equations of motion.
\begin{corollary}\label{coro_ZC}
\begin{eqnarray}
&&\partial_{T}\ \ell(j,\mu)=\MM(j+1,\mu)\ell(j,\mu) - \ell(j,\mu) \MM(j,\mu)\\
&&  \partial_{T}\ k^-(\mu)=\MM(1,\mu)\ k^-(\mu) - k^-(\mu)\ \MM(1,-\mu)\\
&&   \partial_{T}\ k^+(\mu)=\MM(N+1,-\mu)\ k^+(\mu) - k^+(\mu)\ \MM(N+1,\mu)
 \end{eqnarray}
 \end{corollary}

 \subsection{Discussion}
 
In the non-dynamical case \textit{i.e.} when $k^\pm$ satisfy
\begin{equation}\label{eq:ndRE}
 \{k^\pm(\lda),k^\pm(\mu)\}=0\,,
\end{equation}
we find that $\{ \cb(\lambda)\ ,\ k_b^\pm(\mu) \}=0$ and consequently, 
 \begin{eqnarray}
 &&\MM(1,\mu) k^-(u)=k^-(u) \MM(1,-\mu)\\
 &&\MM(N+1,-\mu) k^+(u)=k^+(u) \MM(N+1,\mu)\;.
\end{eqnarray}
These look formally like \eqref{relation_KM} used in \cite{Sk} but with one crucial difference; they involve the double-row matrices $\MM(1,\mu)$, $\MM(N+1,\mu)$ and not the single-row matrix $M(j,\mu)$ evaluated at the end points. They are different in general. This will be clear on the Toda chain used as an example below. The use of the single-row matrix in \eqref{relation_KM} and its dynamical variant has been at the origin of the various discrepancies raised in the introduction. Our results show that no such problems arise if we use the correct $M$ matrix for the Lax pair formulation of the system at hand. The latter should be derived from the boundary version of the Semenov-Tian-Shansky formula. 

From the Hamiltonian point of view, the previous discussion amounts to saying that the time flow $T$ on the bulk fields of the system induced by the double-row transfer matrix is different in general from the time flow $t$ induced by the single-row transfer matrix. Only in very special cases (when the solutions of the reflection equation and of \eqref{relation_KM} coincide) can one reconcile the two flows by imposing boundary conditions on the fields that are dictated by \eqref{relation_KM}. In retrospect, that this ``extrinsic approach''\footnote{By this we mean that boundary conditions are formulated on the bulk fields.} was possible at all is rather remarkable. In our approach, which we could call ``intrinsic'', no boundary conditions are explicitely imposed on the bulk fields. Instead, the coupling between bulk and boundary fields is governed by our zero-curvature equations. It is sometimes possible to go from one approach to the other, as will be illustrated on the Toda chain below, and the distinction becomes irrelevant. 

\section{Hamiltonian and zero-curvature presentations for models with open boundaries: Continuous case}\label{Sec2}

For continuous models described by a Lax matrix $\ell(x,\lambda)$ our starting point is the following 
ultralocal Poisson algebra
\begin{equation}
\{\ell_a(x,\lambda)\ ,\ \ell_b(y,\mu)\}=\delta(x-y)\left[ r_{ab}(\lambda-\mu)\ , \ \ell_a(x,\lambda) +\ell_b(y,\mu) \right]\,,
\end{equation} 
which we complement with the following boundary Poisson algebra
\begin{eqnarray}
\{k_a^-(\lambda)\ ,\ k_b^-(\mu) \}&=&r_{ab}(\lambda-\mu)k_a^-(\lambda) k_b^-(\mu) -   k_a^-(\lambda)   \ k_b^-(\mu) r_{ba}(\lambda-\mu)\nonumber\\
&&+k_a^-(\lambda)r_{ba}(\lambda+\mu) k_b^-(\mu)-k_b^-(\mu)r_{ab}(\lambda+\mu) k_a^-(\lambda)\,,\\
\{k_a^+(\lambda)\ ,\ k_b^+(\mu) \}&=&r_{ba}(\lambda-\mu)k_a^+(\lambda) k_b^+(\mu) -   k_a^+(\lambda)   \ k_b^+(\mu) r_{ab}(\lambda-\mu)\nonumber\\
&&+k_a^+(\lambda)r_{ab}(\lambda+\mu) k_b^+(\mu)-k_b^+(\mu)r_{ba}(\lambda+\mu) k_a^+(\lambda)\,,\\
\{k^-_a(\lambda)\ ,\ k^+_b(\mu) \}&=&0\,,\\
\{k^\pm_a(\lambda)\ ,\ \ell_b(x,\mu)\}&=&0\;.
\end{eqnarray}

Associated to $\ell(x, \lda)$ is the transition matrix $T(x,y,\lda)$, $y<x$, defined by
\begin{equation}
\partial_xT(x,y,\lda)=\ell(x,\lda)\ T(x,y,\lda)\,,~~T(x,y,\lda)|_{x=y}=\1\,.
\end{equation}
 It satisfies
 \begin{eqnarray}
\{T_a(x,y,\lda)\ ,\ T_b(x,y,\mu)\}=\left[ r_{ab}(\lambda-\mu)\ , \ T_a(x,y,\lda)\ T_b(x,y,\mu) \right]\,.
 \end{eqnarray}
 
 The continuous double-row transfer matrix on the interval $[0,L]$ is defined by
 \begin{eqnarray}\label{drtr_cont}
 \cb(\lambda)=tr_a\big(\ k^+_a(\lambda) \ T_a(\lambda) \  k^-_a(\lambda)\  T_a(-\lambda)^{-1}\ \big)
 \end{eqnarray}
 where $T(\lda)=T(L,0,\lda)$, and satisfies 
  \begin{eqnarray}\label{drtr_com}
  \{\cb(\lambda),\cb(\mu) \}=0\,.
  \end{eqnarray}
  Following the construction of \cite{AD1} which we generalise to the present case of dynamical $k$ matrices, we define, for $x\in[0,L]$, 
 \begin{eqnarray}\label{eq:Mad_cont}
 \MM_b(x,\lambda,\mu)&=&tr_a\big(\ k^+_a(\lambda) \ T_a(L,x,\lambda)\  r_{ab}(\lambda-\mu)\ T_a(x,0,\lambda) \ k^-_a(\lambda)\ T_a(-\lambda)^{-1}\ \big)\nonumber\\
 &+&tr_a\big(\ k^+_a(\lambda)  \ T_a(\lambda)\  \ k^-_a(\lambda)\ T_a(x,0,-\lambda)^{-1}\ r_{ba}(\lambda+\mu)\ T_a(L,x,-\lambda)^{-1}\ \big)
 \end{eqnarray}
 Then, similarly to the discrete case, we prove
 \begin{thm}
 	\begin{eqnarray}
\{ \cb(\lambda)\ , \ \ell_b(x,\mu) \}&=&\partial_x\MM_b(x,\lda,\mu)+[\MM_b(x,\lambda,\mu), \ell_b(x,\mu)]\\
\{ \cb(\lambda)\ ,\ k_b^-(\mu) \}&=& \MM_b(0,\lambda,\mu)\ k_b^-(\mu) - k_b^-(\mu)\ \MM_b(0,\lambda,-\mu)\\
\{ \cb(\lambda)\ ,\ k_b^+(\mu) \}&=&\MM_b(L,\lambda,-\mu)\ k_b^+(\mu) - k_b^+(\mu)\ \MM_b(L,\lambda,\mu)
\end{eqnarray}
 \end{thm}
Denoting $\MM(x,\mu)$ the matrix extracted from $\MM(x,\lambda,\mu)$ consistently with an extraction of 
 $\HH$ from $\cb(\lda)$, we obtain the zero curvature representation of the equations of motion.
 \begin{corollary}
 	\begin{eqnarray}
 	&&\partial_{T}\ \ell(x,\mu)=\partial_x\MM(x,\mu) + [ \MM(x,\mu),\ell(x,\mu)]\\
 	&&  \partial_{T}\ k^-(\mu)=\MM(0,\mu)\ k^-(\mu) - k^-(\mu)\ \MM(0,-\mu)\\
 	&&   \partial_{T}\ k^+(\mu)=\MM(L,-\mu)\ k^+(\mu) - k^+(\mu)\ \MM(L,\mu)
 	\end{eqnarray}
 \end{corollary}

\section{Example: the finite Toda chain revisited}\label{Sec3}

\subsection{Non dynamical boundaries and $BC_N$ Toda lattice}

We consider the Toda chain with coordinates $\{ x_j\ |\ j=1,\dots,N\}$ and canonical momenta $\{X_j\ |\ j=1,\dots,N \}$ satisfying
\begin{equation}\label{eq:symp}
 \{x_j\ , \ x_k \}=\{X_j\ , \ X_k \}=0\quad\text{and}\qquad \{X_j\ , \ x_k \}=\delta_{jk}\;.
\end{equation}
The Lax matrix 
\begin{eqnarray}
\label{ell_Toda}
 \ell(j,u)=\begin{pmatrix}
         u+X_j& -e^{x_j}\\
         e^{-x_j} & 0
        \end{pmatrix}
\end{eqnarray}
satisfies relation \eqref{rll} with $r$ being the rational classical $r$-matrix : 
\begin{equation}
r_{12}(\lda)=\frac{P_{12}}{\lda}\,.
\end{equation}
The most general solutions (up to an irrelevant overall function of $\lda$) of the non-dynamical reflection equations (\eqref{rkk1}-\eqref{rkk2} with $\{k_a^\pm(\lambda)\ ,\ k_b^\pm(\mu) \}=0$) are
\begin{eqnarray}
 k^-(\lda)=\begin{pmatrix}
        \lda\theta_1 +\alpha_1 & \lda\\
         -\beta_1 \lda & -\lda\theta_1 +\alpha_1
        \end{pmatrix}\quad\text{and}\qquad
 k^+(\lda)=\begin{pmatrix}
        \lda\theta_N +\alpha_N & \lda\beta_N\\
         - \lda & -\lda\theta_N +\alpha_N
        \end{pmatrix}  \;,     
\end{eqnarray}
where $\alpha_1,\ \beta_1,\ \theta_1,\ \alpha_N,\ \beta_N$ and $\theta_N$ are arbitrary parameters.
In this case,  we get the following Hamiltonian from the coefficient in front of $\lambda^{2N}$ (multiplied by $(-1)^N/2$) of the double-row transfer matrix \eqref{drtr},
\begin{equation}\label{eq:H}
 \HH= \sum_{j=1}^N \frac{1}{2} X_j^2 +\sum_{j=1}^{N-1} e^{x_{j+1}-x_j} +B^-(x_1,X_1) + B^+(x_N,X_N)
\end{equation}
with
\begin{eqnarray}
 B^-(x_1,X_1)&=&\alpha_1 e^{x_1}+\frac{\beta_1}{2}e^{2x_1} + \theta_1 X_1 e^{x_1}\quad\text{,}\quad
  B^+(x_N,X_N)=\alpha_N e^{-x_N}+\frac{\beta_N}{2}e^{-2x_N} + \theta_N X_N e^{-x_N}.\quad
\end{eqnarray}
To the best of our knowledge, this is the first time that these boundary conditions are presented but as we show below, they produce equations of motion that are canonically equivalent to the known case $\theta_1=\theta_N=0$.
For $\theta_1=\theta_N=0$, this Hamiltonian becomes the one studied in \cite{Sk,Ino}. For $\theta_1=\theta_N=\alpha_1=\alpha_N=0$, we recover the Toda chain introduced in \cite{Bog}. The associated Hamilton's equations of motion are
\begin{alignat}{3}
& \dot{x_j}=X_j\quad,\quad &&\dot{X_j}=e^{x_{j+1}-x_j}-e^{x_{j}-x_{j-1}}\ , \qquad\text{for}\qquad j=2,\dots,N-1\label{eq:eom1}\\
& \dot{x_1}=X_1+\theta_1 e^{x_1}\quad,\quad && \dot{X_1}=e^{x_{2}-x_1}-\alpha_1 e^{x_1}-\beta_1 e^{2x_1}-\theta_1 X_1 e^{x_1}\ ,\label{eq:eom2}\\
& \dot{x_N}=X_N+\theta_N e^{-x_N}\quad,\quad && \dot{X_N}=-e^{x_{N}-x_{N-1}}+\alpha_N e^{-x_N}+\beta_N e^{-2x_N}+\theta_N X_N e^{-x_N}\,,\label{eq:eom3}
\end{alignat}
where the dot indicates differentiation with respect to $T$ associated to $\HH$ in \eqref{eq:H}.
Now, from relation \eqref{eq:Mad}, we can derive the second part of the Lax pair 
\begin{eqnarray}
 \MM(j,\mu)&=&\begin{pmatrix}
        \displaystyle -\frac{\mu}{2} & e^{x_j}\\
        -e^{x_{j-1}} &\displaystyle \frac{\mu}{2}
        \end{pmatrix}\qquad\text{for}\qquad j=2,\dots,N\\
\MM(1,\mu)&=&\begin{pmatrix}
        \displaystyle -\frac{\mu}{2}+\theta_1 e^{x_1} &e^{x_1} \\
         \mu\theta_1 -\alpha_1-\beta_1e^{x_1}& \displaystyle\frac{\mu}{2}-\theta_1 e^{x_1}
        \end{pmatrix}\\
\MM(N+1,\mu)&=&\begin{pmatrix}
        \displaystyle  -\frac{\mu}{2} +\theta_N e^{-x_N}&-\mu\theta_N+\alpha_N+\beta_N e^{-x_N} \\
       -e^{-x_N}  &  \displaystyle\frac{\mu}{2}-\theta_N e^{-x_N}
        \end{pmatrix}\;.
\end{eqnarray}
We recover the matrices given in \cite{kS} by adding the irrelevant term $-\mu/2\1$ and using the canonical change of variables $\tilde{x}_j=x_j$, $j=1,\dots,n$, $\tilde{X}_j=X_j$, $j=2,\dots,n-1$, $\tilde{X}_1=X_1+\theta_1e^{x_1}$ and $\tilde{X}_n=X_n+\theta_ne^{-x_n}$.
As stated in the main theorem, the Hamilton's equations of motion \eqref{eq:eom1}-\eqref{eq:eom3} are equivalent to the following zero curvature equation, as can be checked directly,
\begin{eqnarray}
&&\dot \ell(j,\mu) =\MM(j+1,\mu) \ell_j(\mu) -\ell_j(\mu)\MM(j,\mu)\,,~~j=1,\dots,N\,,
\end{eqnarray}
while the boundary equations
\begin{eqnarray}
&&\MM(1,\mu) k^-(\mu)=k^-(\mu) \MM(1,-\mu)\,,\\
&&\MM(N+1,-\mu) k^+(\mu)=k^+(\mu) \MM(N+1,\mu)\;,
\end{eqnarray}
are trivially satisfied. 

\subsection{An example of dynamical boundaries and $D_N$ Toda lattice}

In this subsection, we consider an example of a dynamical boundary for the Toda chain. We restrict ourselves to the case where $k^-(\lda)$ is dynamical and $k^+(\lda)$ is non-dynamical and is given by $k^+=\begin{pmatrix} 0 & 0\\ -1 & 0  \end{pmatrix}$.
Let us extend the symplectic space \eqref{eq:symp} by adding $3$ generators $E,F,H$ Poisson commuting with $x_j$ and $X_j$ and satisfying the $\textsl{sl}(2)$ Poisson algebra
\begin{eqnarray}
\{ H, E\}=E\quad,\qquad \{ H, F\}=-F\quad,\qquad \{ E, F\}=2H\,.
\end{eqnarray}
Let us recall that there is a Casimir $C=H^2+EF$ which we set to the value $c_1/4$. Then
\begin{eqnarray}
k^-(\lda)=\begin{pmatrix}
\lda/2-H & F\\
F & \lda/2+H
\end{pmatrix}
\end{eqnarray}
satisfies the dynamical reflection equation \eqref{rkk1}.

In this case,  we get the Hamiltonian from the expansion of the double-row transfer matrix \eqref{drtr} 
\begin{equation}
\cb(\lda)=\lda^{2N}\HH^{(2N)}+\lda^{2N-2} \HH^{(2N-2)}+\dots
\end{equation}
as $\HH=-\frac{1}{2}\frac{\HH^{(2N-2)}}{\HH^{(2N)}}$, yielding
\begin{equation}\label{eq:Hdyn}
\HH= \sum_{j=1}^N \frac{1}{2} X_j^2 +\sum_{j=1}^{N-1} e^{x_{j+1}-x_j} +B
\end{equation}
with
\begin{eqnarray}
B&=&\frac{1}{2(F-e^{x_1})}\left(e^{x_2}+X_1^2 e^{x_1}-2H e^{x_1}X_1-Ee^{2x_1}\right).
\end{eqnarray}
From this Hamiltonian, we can compute the equations of motion. In particular, we can show that
$ \dot F =\dot x_1 e^{x_1}$ which we can integrate to $F=e^{x_1}+c_0/2$, with $c_0$ a constant, and eliminate $F$.
It is convenient to use the following change of coordinate
\begin{equation}
e^{\tilde x_1}= \frac{c_0 e^{x_1}}{c_0+e^{x_1}}
\end{equation}
to write down the equations of motion. After some algebraic manipulations, they read
\begin{alignat}{3}
& \ddot{x}_j=e^{x_{j+1}-x_j}-e^{x_{j}-x_{j-1}}\ , \qquad\text{for}\qquad j=3,\dots,N-1\label{eq:eomyd1}\\
& \ddot{x}_2=e^{x_{3}-x_2}-e^{x_{2}-\tilde x_{1}}\ , \label{eq:eomyd1b}\\
& \ddot{\tilde x}_1=e^{x_{2}-\tilde x_1}-e^{\tilde x_1-x_0} ,\label{eq:eomdy2}\\
& \ddot{x}_N=-e^{x_{N}-x_{N-1}},\label{eq:eomdyn3}
\end{alignat}
where we have introduced a coordinate $x_0$ defined by
\begin{equation}
\label{BC_x0}
e^{-x_0}=\frac{e^{x_2}}{c_0^2}+\frac{((\dot{\tilde x}_1)^2-c_1)e^{x_1}}{c_0^2-e^{2\tilde x_1}}\;.
\end{equation}
Here, the dot indicates differentiation with respect to $T$ associated to $\HH$ in \eqref{eq:Hdyn}.
The use of $x_0$ is a realisation of the link between the intrinsic and extrinsic approach to boundary conditions mentiond in the discussion above. The equations of motion could have been written without resorting to an extra variable $x_0$ (intrinsic viewpoint) and all the effect of the boundary would have appeared in extra terms in the equation for ${\tilde x}_1$. By introducing $x_0$, we see that the equation for ${\tilde x}_1$ takes the same bulk form as the equation for $x_j$, $j=3,\dots,N-1$ and \eqref{BC_x0} plays now the role of a boundary condition (extrinsic viewpoint). 

We recognize in \eqref{BC_x0} a particular case of the boundary condition introduced in \cite{AH}. As shown in \cite{AH}, this model can be seen as 
a generalization of the $D_N$-Toda lattice. The double-row construction for this Hamiltonian was also considered in \cite{KT}.

With our constructions, we are now able to derive the second part of the Lax pair from relation \eqref{eq:Mad},
\begin{eqnarray}
\MM(j,\mu)&=&\begin{pmatrix}
\displaystyle -\frac{\mu}{2} & e^{x_j}\\
-e^{x_{j-1}} &\displaystyle \frac{\mu}{2}
\end{pmatrix}\qquad\text{for}\qquad j=3,\dots,N\\
\MM(2,\mu)&=&\begin{pmatrix}
\displaystyle -\frac{\mu}{2} & e^{x_2}\\
\frac{e^{x_1}-2F}{2e^{x_1}(F-e^{x_1})} &\displaystyle \frac{\mu}{2}
\end{pmatrix}\\
\MM(N+1,\mu)&=&\begin{pmatrix}
\displaystyle  -\frac{\mu}{2}&0 \\
-e^{-x_N}  &  \displaystyle\frac{\mu}{2}
\end{pmatrix}\;,
\end{eqnarray}
and 
\begin{eqnarray}
\MM(1,\mu)= \frac{1}{2(e^{x_1}-F)}\begin{pmatrix}
\displaystyle \mu F +e^{x_1}(2H-X_1) &e^{x_1}(e^{x_1}-2F) \\
\mu^2+2\mu H +\frac{e^{x_2}+X_1^2 e^{x_1}-2H e^{x_1}X_1-Ee^{2x_1}-2EFe^{x_1}}{F-e^{x_1}}& -\mu F -e^{x_1}(2H-X_1)
\end{pmatrix}\;.
\end{eqnarray}
As stated in the corollary \ref{coro_ZC}, the equations of motion can be obtained equivalently as a zero-curvature condition
using the above Lax matrices $\MM(j,\mu)$ and $\ell(j,\lda)$ in \eqref{ell_Toda}. In particular, the equations of motion of $E,F$ and $H$ are obtained from
\begin{eqnarray}
&&  \dot{k}^-(\mu)=\MM(1,\mu)\ k^-(\mu) - k^-(\mu)\ \MM(1,-\mu).
\end{eqnarray}

\section*{Perspectives}\label{Sec4}

We have established here a complete picture for a description of the zero-curvature formulation (Lax formulation) of a $1+1$ dimensional integrable system on the interval, starting from a Lax matrix (i.e. the space
component of the flat connection) boundary matrices and their Poisson structure. The latter is assumed to be ultralocal and parametrized by a skew-symmetric non dynamical $r$-matrix. The boundary matrices obey a dynamical quadratic Poisson structure of soliton-preserving type.

Relaxing these restrictions may lead to formulating (Lax) zero-curvature conditions for more general $1+1$ integrable systems with boundaries. The more obvious procedure is to consider so-called ``soliton 
non-preserving conditions'' where the $r$ matrices parametrizing the Poisson structure of the $k$ matrices are more generic, as in e.g. \cite{AD1}, the most general case being of course \cite{FrMa} where no 
connection is now assumed between ``outside'' and ``inside'' matrices in the quadratic Poisson structure \eqref{Kdyn1}-\eqref{Kdyn2}. In any case, it is a priori required by the coaction properties of 
the quadratic Poisson algebra \cite{FrMa} that the bulk single-row monodromy matrices $L$ acting on $k$ to generalize the double-row transfer matrix \eqref{drtr} still obey a purely Sklyanin-type quadratic 
Poisson structure with a single skew-symmetric $r$ matrix, typically one of the ``outside'' matrices. Such a structure is indeed identified for bulk single-row monodromy matrices even when derived from Lax
matrices with non skew-symmetric, non ultralocal Poisson structures (see \cite{Ma}). Typical examples of such objects are theories on symmetric spaces for which a wealth of non skew symmetric $r$ matrices or $r,s$ pairs are 
available (see e.g. \cite{AT}).

The next step is to consider the case of Lax matrices with dynamical $r$-matrices depending on the field variables, and therefore not obeying the classical Yang-Baxter equation but some adequately modified version of it \cite{Ma}. A typical case is the complex sine-Gordon model. The issue is then to find actual representations of the corresponding Poisson algebra to get consistent matrix $k^\pm$. Since now the $r$-matrix depends on the bulk fields at the limit point so would also $k^\pm$, hence the statement \eqref{kloc3} that $k^\pm$ has zero Poisson brackets with the bulk fields may fail and so does our whole construction. Moreover in the case of cSG even the exact formulation of a consistent Poisson structure for $k$ matrices derived from such dynamical $r$-matrices is lacking. This situation clearly requires some fundamental reconsideration.
For other dynamical $r$-matrices such as arising in Calogero-Moser \cite{ABT} and Ruijsenaar-Schneider models \cite{Suris}, the situation regarding the reflection algebra is much better understood \cite{ACF,ANR} but this time, non-trivial field-theoretical Lax matrices obeying the co-action Poisson structure are lacking.

\paragraph{Acknowledgment:} V. Caudrelier and N. Cramp\'e wish to thank the LPTM, Universit\'e Cergy-Pontoise and Institut d'\'Etudes Avanc\'ees, UCP where this work was initiated. J. Avan and N. Cramp\'e acknowledge the hospitality 
of the School of Mathematics, University of Leeds where this work was completed. N. Cramp\'e's visit was partially supported by the Research Visitor Centre of the School of Mathematics.

\end{document}